\RequirePackage[2020-02-02]{latexrelease}

\documentclass[twocolumn,preprintnumbers]{revtex4}%
\usepackage{amssymb}
\usepackage{amsmath}
\usepackage{graphicx}
\usepackage{epstopdf}
\usepackage{dcolumn}
\usepackage{bm}
\usepackage{xcolor}
\usepackage{ifthen}
\usepackage[normalem]{ulem}
\usepackage{amsfonts}%
\RequirePackage[2020-02-02]{latexrelease}
\RequirePackage[2020-02-02]{latexrelease}
\RequirePackage[2020-02-02]{latexrelease}
\RequirePackage[2020-02-02]{latexrelease}
\UseRawInputEncoding
\providecommand{\U}[1]{\protect\rule{.1in}{.1in}}
\providecommand{\U}[1]{\protect\rule{.1in}{.1in}}
\def\showal{1}
\newcommand{\al}[1]{\ifthenelse{\showal=1}{\textcolor{orange}{[[#1]]}}{}}

\newcommand{\eb}[1]{\ifthenelse{\showal=1}{\textcolor{cyan}{[[#1]]}}{}}
\begin{document}
\title{Spontaneous disentanglement of indistinguishable particles}
\author{Eyal Buks}
\email{eyal@ee.technion.ac.il}
\affiliation{Andrew and Erna Viterbi Department of Electrical Engineering, Technion, Haifa
32000, Israel}
\date{\today }

\begin{abstract}
A master equation containing a nonlinear term that gives rise to
disentanglement has been recently investigated. In this study, a modified
version, which is applicable for indistinguishable particles, is proposed, and
explored for both the Bose-Hubbard and the Fermi-Hubbard models. It is found
for both Bosons and Fermions that disentanglement can give rise to quantum
phase transitions.

\end{abstract}
\maketitle

%Force line breaks with \\

%Lines break automatically or can be forced with \\

%It is always \today, today,
%but any date may be explicitly specified

%PACS, the Physics and Astronomy
%Classification Scheme.
%\keywords{Suggested keywords}%Use showkeys class option if keyword
%display desired

\textbf{Introduction} - In standard quantum mechanics (QM), the time evolution
of the density operator $\rho$ is governed by a master equation that has a
linear dependency on $\rho$ \cite{Lindblad_119}. A variety of nonlinear
extensions to QM have been proposed
\cite{Weinberg_61,Doebner_3764,Gisin_5677,Gisin_2259,Kaplan_055002,Munoz_110503,Jacobs_279,Geller_2200156}%
. These proposals have been mainly motivated by the old-standing problem of
quantum measurement \cite{Schrodinger_807}. For some cases, the proposed
extension can give rise to state vector spontaneous collapse
\cite{Bassi_471,Pearle_857,Ghirardi_470,Bassi_257,Bennett_170502,Kowalski_1,Fernengel_385701,Kowalski_167955,Oppenheim_041040,Schrinski_133604}%
. Another motivation is related to the observation that multistability in
finite systems cannot be derived from a master equation having linear
dependency on $\rho$. This observation raises the question of how
experimentally-observed processes such as phase transitions
\cite{Chomaz_68,mainwood2005phase,Callender_539,Liu_S92,Ardourel_99,Shech_1170,toda1978statistical_I,Sakthivadivel_035}
and dynamical instabilities \cite{Suhl_209} can be accounted for within the
framework of standard QM \cite{Buks_012439}. For some cases, nonlinearity may
give rise to conflicts with well-established physical principles, such as
causality
\cite{Bassi_055027,Jordan_022101,Polchinski_397,Helou_012021,Rembielinski_012027,Rembielinski_420}
and separability \cite{Hejlesen_thesis,Jordan_022101,Jordan_012010}. In
addition, some predictions of standard QM, which have been experimentally
confirmed to very high accuracy, are inconsistent with some of the proposed
nonlinear extensions.

A nonlinear mechanism giving rise to suppression of entanglement (i.e.
disentanglement) has been recently proposed \cite{Buks_2400036}. This
mechanism of disentanglement, which makes the collapse postulate redundant, is
introduced by adding a nonlinear term to the master equation [see Eq.
(\ref{MME}) below]. The proposed modified master equation can be constructed
for any physical system whose Hilbert space has finite dimensionality, and it
does not violate norm conservation of the time evolution. The nonlinear term
added to the master equation has no effect on any product (i.e. disentangled)
state. For a multipartite system, disentanglement between any pair of
subsystems can be introduced by the added nonlinear term. Moreover,
thermalization can be incorporated by an additional nonlinear term added to
the master equation. The nonlinear extension that was proposed in
\cite{Buks_2400036} has been explored for systems containing distinguishable
particles. The current work is devoted to disentanglement of indistinguishable
particles. A modified mater equation having a nonlinear term that generates
disentanglement of indistinguishable particles is proposed. The dynamics
generated by the proposed master equation is explored for both the
Boson-Hubbard and the Fermi-Hubbard models
\cite{Hubbard_238,Hubbard_401,Gutzwiller_159,Kanamori_275}.

\textbf{Indistinguishable particles} - The Hilbert space of a system
containing indistinguishable particles can be constructed using a given
orthonormal basis $\left\{  \left\vert a_{i}\right\rangle \right\}  _{i}$ that
spans the single-particle Hilbert space \cite{fetter2012quantum}. The ket
vector $\left\vert \bar{n}\right\rangle $ represents a state that is
characterized by the vector $\bar{n}=\left(  n_{1},n_{2},\cdots\right)  $,
where the integer $n_{i}$ is the number of particles that occupy the single
particle state $\left\vert a_{i}\right\rangle $. The set $\left\{  \left\vert
\bar{n}\right\rangle \right\}  _{\bar{n}}$ forms an orthonormal basis for the
many-particle Hilbert space. The state $\left\vert \bar{n}\right\rangle $ can
be expressed in terms of the creation operators $a_{i}^{\dag}$ as $\left\vert
\bar{n}\right\rangle =\left(  n_{1}!n_{2}!\cdots\right)  ^{-1/2}\left(
a_{1}^{\dag}\right)  ^{n_{1}}\left(  a_{2}^{\dag}\right)  ^{n_{2}}%
\cdots\left\vert 0\right\rangle $, where $\left\vert 0\right\rangle $
represents the state where all occupation numbers are zero. The number
operator $N_{i}$ is defined by $N_{i}=a_{i}^{\dag}a_{i}$.

It is postulated that the following commutation relations hold $\left[
a_{i^{\prime}}^{{}},a_{i^{\prime\prime}}^{{}}\right]  _{+}=\left[
a_{i^{\prime}}^{\dag},a_{i^{\prime\prime}}^{\dag}\right]  _{+}=0$ and $\left[
a_{i^{\prime}}^{{}},a_{i^{\prime\prime}}^{\dag}\right]  _{+}=\delta
_{i^{\prime},i^{\prime\prime}}$ ($\left[  a_{i^{\prime}}^{{}},a_{i^{\prime
\prime}}^{{}}\right]  _{-}=\left[  a_{i^{\prime}}^{\dag},a_{i^{\prime\prime}%
}^{\dag}\right]  _{-}=0$ and $\left[  a_{i^{\prime}}^{{}},a_{i^{\prime\prime}%
}^{\dag}\right]  _{-}=\delta_{i^{\prime},i^{\prime\prime}}$) for Fermions
(Bosons), where $\left[  A,B\right]  _{\pm}=AB\pm BA$ for general operators
$A$ and $B$. For both Fermions and Bosons $N_{i}\left\vert \bar{n}%
\right\rangle =n_{i}\left\vert \bar{n}\right\rangle $. The relation
$N_{i}=N_{i}^{2}$, which holds for Fermions, yields the Pauli's exclusion
principle. A single-particle unitary transformation mapping from the
orthonormal basis $\left\{  \left\vert a_{i}\right\rangle \right\}  _{i}$ to
an alternative orthonormal basis $\left\{  \left\vert b_{j}\right\rangle
\right\}  _{j}$ yields the many-particle creation operator transformations
given by $b_{j}^{\dag}=\sum_{i}\left\langle a_{i}\right.  \left\vert
b_{j}\right\rangle a_{i}^{\dag}$, where $\left\langle a_{i}\right.  \left\vert
b_{j}\right\rangle $ is the inner product of the single-particle states
$\left\vert a_{i}\right\rangle $ and $\left\vert b_{j}\right\rangle $.

Observables of a system of identical particles must be defined in a way that
is consistent with the principle of indistinguishability. Consider two-body
interaction that is represented by an Hermitian operator $V_{\mathrm{TP}}$ on
the two-particle Hilbert space. A basis for this Hilbert space can be
constructed using a given orthonormal basis for the single-particle Hilbert
space $\left\{  \left\vert b_{j}\right\rangle \right\}  _{j}$. When the two
particles are considered as distinguishable, the basis of the two-particle
Hilbert space can be taken to be $\left\{  \left\vert j^{\prime}%
,j^{\prime\prime}\right\rangle \right\}  _{j^{\prime},j^{\prime\prime}}$. The
ket vector $\left\vert j^{\prime},j^{\prime\prime}\right\rangle $ represents a
state, for which the first particle is in single particle state $\left\vert
b_{j^{\prime}}\right\rangle $, and the second one is in state $\left\vert
b_{j^{\prime\prime}}\right\rangle $. Assume the case where the single-particle
basis vectors $\left\vert b_{j}\right\rangle $ are chosen in such a way that
diagonalizes $V_{\mathrm{TP}}$, i.e. $V_{\mathrm{TP}}\left\vert j^{\prime
},j^{\prime\prime}\right\rangle =v_{j^{\prime},j^{\prime\prime}}\left\vert
j^{\prime},j^{\prime\prime}\right\rangle $, where the eigenvalue
$v_{j^{\prime},j^{\prime\prime}}$ is given by $v_{j^{\prime},j^{\prime\prime}%
}=\left\langle j^{\prime},j^{\prime\prime}\right\vert V_{\mathrm{TP}%
}\left\vert j^{\prime},j^{\prime\prime}\right\rangle $. In the many-particle
case, the two-particle interaction $V_{\mathrm{TP}}$ is represented by the
operator $V$, which for both Fermions and Bosons is given by%
\begin{equation}
V=\frac{1}{2}\sum_{j^{\prime},j^{\prime\prime}}v_{j^{\prime},j^{\prime\prime}%
}N_{j^{\prime}}\left(  N_{j^{\prime\prime}}-\delta_{j^{\prime},j^{\prime
\prime}}\right)  \ , \label{V TP}%
\end{equation}
where $N_{j}=b_{j}^{\dag}b_{j}^{{}}$. Note that the matrix element
$\left\langle \bar{n}\right\vert V\left\vert \bar{n}\right\rangle $ is given
by $\left\langle \bar{n}\right\vert V\left\vert \bar{n}\right\rangle
=\sum_{j^{\prime}<j^{\prime\prime}}n_{j^{\prime}}n_{j^{\prime\prime}%
}v_{j^{\prime},j^{\prime\prime}}+\left(  1/2\right)  \sum_{j}n_{j}\left(
n_{j}-1\right)  v_{j,j}$. While the factor $n_{j^{\prime}}n_{j^{\prime\prime}%
}$ represents the number of particle pairs occupying single-particle states
$j^{\prime}$ and $j^{\prime\prime}$ for the case $j^{\prime}\neq
j^{\prime\prime}$ , the factor $n_{j}\left(  n_{j}-1\right)  /2$ represents
the number of particle pairs occupying the same single-particle state $j$. The
expression given by Eq. (\ref{V TP}) for the two-particle interaction in the
basis that diagonalizes $V_{\mathrm{TP}}$, is employed below for the
construction of the nonlinear disentanglement terms added to the master equation.

\textbf{Master equation} - The modified master equation for the density
operator $\rho$ takes a form given by
\cite{Kaplan_055002,Geller_2200156,Grimaudo_033835,Kowalski_167955,Buks_2400036}%
\begin{equation}
\frac{\mathrm{d}\rho}{\mathrm{d}t}=i\hbar^{-1}\left[  \rho,\mathcal{H}\right]
-\Theta\rho-\rho\Theta+2\left\langle \Theta\right\rangle \rho\;, \label{MME}%
\end{equation}
where $\hbar$ is the Planck's constant, $\mathcal{H}^{{}}=\mathcal{H}^{\dag}$
is the Hamiltonian, the operator $\Theta^{{}}=\Theta^{\dag}$ is allowed to
depend on $\rho$, and $\left\langle \Theta\right\rangle =\operatorname{Tr}%
\left(  \Theta\rho\right)  $. Note that $\mathrm{d}\operatorname{Tr}%
\rho/\mathrm{d}t=0$ provided that $\operatorname{Tr}\rho=1$ (i.e. $\rho$ is
normalized), and that $\mathrm{d}\operatorname{Tr}\rho^{2}/\mathrm{d}t=0$,
provided that $\rho^{2}=\rho$ (i.e. $\rho$ represents a pure state).

For the case $\mathcal{H}=0$, and for a fixed operator $\Theta$, the
expectation value $\left\langle \Theta\right\rangle $ monotonically decreases
with time. Hence, the nonlinear term in the modified master equation
(\ref{MME}) can be employed to suppress a given physical property, provided
that $\left\langle \Theta\right\rangle $ quantifies that property. The
operator $\Theta$ is taken to be given by $\Theta=\gamma_{\mathrm{H}%
}\mathcal{Q}^{\left(  \mathrm{H}\right)  }+\gamma_{\mathrm{D}}\mathcal{Q}%
^{\left(  \mathrm{D}\right)  }$, where both rates $\gamma_{\mathrm{H}}$ and
$\gamma_{\mathrm{D}}$ are positive, and both operators $\mathcal{Q}^{\left(
\mathrm{H}\right)  }$ and $\mathcal{Q}^{\left(  \mathrm{D}\right)  }$\ are
Hermitian. The first term $\gamma_{\mathrm{H}}\mathcal{Q}^{\left(
\mathrm{H}\right)  }$ gives rise to thermalization
\cite{Grabert_161,Ottinger_052119}, whereas disentanglement is generated by
the second term $\gamma_{\mathrm{D}}\mathcal{Q}^{\left(  \mathrm{D}\right)  }$.

Consider the master equation (\ref{MME}) for the case where $\mathcal{H}^{{}}$
is time independent, $\gamma_{\mathrm{D}}=0$ (i.e. no disentanglement), and
$\mathcal{Q}^{\left(  \mathrm{H}\right)  }=\beta\mathcal{U}_{\mathrm{H}}$,
where $\mathcal{U}_{\mathrm{H}}=\mathcal{H}+\beta^{-1}\log\rho$ is the
Helmholtz free energy operator, $\beta=1/\left(  k_{\mathrm{B}}T\right)  $ is
the thermal energy inverse, $k_{\mathrm{B}}$ is the Boltzmann's constant, and
$T$ is the temperature. For this case, the thermal equilibrium density matrix
$\rho_{0}$, which is given by $\rho_{0}=e^{-\beta\mathcal{H}}%
/\operatorname{Tr}\left(  e^{-\beta\mathcal{H}}\right)  $, is a fixed-point
steady state solution of the master equation (\ref{MME}), for which the
Helmholtz free energy $\left\langle \mathcal{U}_{\mathrm{H}}\right\rangle $ is
minimized \cite{Jaynes_579,Grabert_161,Ottinger_052119,Buks_052217}. The rate
$\gamma_{\mathrm{H}}$ represents the thermalization inverse time. The
approximation $\left\langle \mathcal{U}_{\mathrm{H}}\right\rangle
=\left\langle \mathcal{H}\right\rangle $ can be employed in the low
temperature limit.

\textbf{Two-particle disentanglement} - The disentanglement operator
$\mathcal{Q}^{\left(  \mathrm{D}\right)  }$ is constructed based on the
many-particle representation given by Eq. (\ref{V TP}) for the two-particle
interaction. Each term in Eq. (\ref{V TP}) having the form $N_{j^{\prime}%
}N_{j^{\prime\prime}}$ contributes to $\mathcal{Q}^{\left(  \mathrm{D}\right)
}$ a term given by $\eta_{j^{\prime},j^{\prime\prime}}Q_{j^{\prime}%
,j^{\prime\prime}}\left\langle Q_{j^{\prime},j^{\prime\prime}}\right\rangle $,
where%
\begin{equation}
Q_{j^{\prime},j^{\prime\prime}}=N_{j^{\prime}}N_{j^{\prime\prime}%
}-\left\langle N_{j^{\prime}}\right\rangle \left\langle N_{j^{\prime\prime}%
}\right\rangle \ . \label{Q_jp,jpp}%
\end{equation}
A conflict with the causality principle
\cite{Bassi_055027,Jordan_022101,Polchinski_397,Helou_012021,Rembielinski_012027,Rembielinski_420}
can be avoided , provided that the coefficients $\eta_{j^{\prime}%
,j^{\prime\prime}}$ (which are allowed to depend on $\rho$) are defined such
that disentanglement generated by the operator $\mathcal{Q}^{\left(
\mathrm{D}\right)  }$ is active only when particles interact. However, for the below-discussed Hubbard model, it is assumed that $\eta_{j^{\prime},j^{\prime\prime}}=1$ (since non-locality is irrelevant for this model with finite number of sites). Note that
$\left\langle \bar{n}\right\vert \mathcal{Q}^{\left(  \mathrm{D}\right)
}\left\vert \bar{n}\right\rangle =0$\ for all basis vectors $\left\vert
\bar{n}\right\rangle $ (provided that the many-particle basis state vectors
$\left\vert \bar{n}\right\rangle $ are constructed based on a single-particle
basis, for which the two-particle interaction $V_{\mathrm{TP}}$ is diagonalized).

The relation $\Theta=\gamma_{\mathrm{H}}\mathcal{Q}^{\left(  \mathrm{H}
\right)  }+\gamma_{\mathrm{D}}\mathcal{Q}^{\left(  \mathrm{D}\right)  }$
together with the ME (\ref{MME}) suggest that disentanglement can be accounted
for by replacing the Helmholtz free energy $\left\langle \mathcal{U}%
_{\mathrm{H}}\right\rangle $ by an effective free energy $\left\langle
\mathcal{U}_{\mathrm{e}}\right\rangle $, which is given by $\left\langle
\mathcal{U}_{\mathrm{e}}\right\rangle =\left\langle \mathcal{U}_{\mathrm{H}%
}\right\rangle +\beta^{-1}\left(  \gamma_{\mathrm{D}}/\gamma_{\mathrm{H}%
}\right)  \left\langle \mathcal{Q}^{\left(  \mathrm{D}\right)  }\right\rangle
$. For cases where the Hamiltonian $\mathcal{H}$ is time-independent, the
effective free energy $\left\langle \mathcal{U}_{\mathrm{e}}\right\rangle $ is
locally minimized for fixed-point steady state solutions of the master
equation (\ref{MME}).

Disentanglement is explored below for the one dimensional Hubbard model
\cite{Lieb_1}. In this model, indistinguishable particles occupy a
one-dimensional array containing $L$ sites. The Hamiltonian $\mathcal{H}$ is
characterized by two real parameters, the nearest neighbor hopping coefficient
$t$, and the on-site interaction coefficient $U$. The array is assumed to have
a ring configuration, for which the first $l=1$ and last $l=L$ sites are
coupled. Some analytical results are derived below for the relatively simple
case of two sites (i.e. $L=2$).

\textbf{Bose-Hubbard model} - For the Bose-Hubbard model
\cite{Links_1591,Zhang_591_,Amico_4298}, the one-dimensional array is occupied
by spinless Bosons. The creation and annihilation operators corresponding to
site $l\in\left\{  1,2,\cdots,L\right\}  $ are denoted by $b_{l}^{\dag}$ and
$b_{l}^{{}}$, respectively. The operators $b_{l}^{\dag}$ and $b_{l}^{{}}$
satisfy Bosonic commutation relations. The Hamiltonian $\mathcal{H}%
_{\mathrm{B}}$ is given by%
\begin{equation}
\mathcal{H}_{\mathrm{B}}=-t\sum_{l=1}^{L}\left(  b_{l}^{\dag}b_{l+1}^{{}%
}+b_{l+1}^{\dag}b_{l}^{{}}\right)  +\frac{U}{2}\sum_{l=1}^{L}N_{l}\left(
N_{l}-1\right)  \ , \label{H BHM}%
\end{equation}
where $N_{l}=b_{l}^{\dag}b_{l}^{{}}$. It is assumed that the array has a ring
configuration, and thus, the last ($l=L$) hopping term $b_{l}^{\dag}%
b_{l+1}^{{}}+b_{l+1}^{\dag}b_{l}^{{}}$ [see Eq. (\ref{H BHM})] is taken to be
given by $b_{L}^{\dag}b_{1}^{{}}+b_{1}^{\dag}b_{L}^{{}}$. Note that the
total\ number operator $N$, which is defined by $N=\sum_{l=1}^{L}N_{l}\ $, is
a constant of the motion.

The linear part of the Hamiltonian $\mathcal{H}_{\mathrm{B}}$ (\ref{H BHM})
can be diagonalized using the transformation $b_{l}^{{}}=L^{-1/2}%
\sum_{l^{\prime}=1}^{L}e^{ik_{l}l^{\prime}}a_{l^{\prime}}^{{}}$, where
$k_{l}=2\pi l/L$. In terms of the Bosonic creation and annihilation operators
$a_{l^{\prime}}^{\dag}$ and $a_{l^{\prime}}^{{}}$, the condition $\left\langle
N_{l}^{2}\right\rangle =\left\langle N_{l}\right\rangle ^{2}$ for all
$l\in\left\{  1,2,\cdots,L\right\}  $ is expressed as $\sum^{\prime
}\left\langle a_{l^{\prime\prime}}^{\dag}a_{l^{\prime}}^{{}}a_{l^{\prime
\prime\prime\prime}}^{\dag}a_{l^{\prime\prime\prime}}^{{}}\right\rangle
=\sum^{\prime}\left\langle a_{l^{\prime\prime}}^{\dag}a_{l^{\prime}}^{{}%
}\right\rangle \left\langle a_{l^{\prime\prime\prime\prime}}^{\dag
}a_{l^{\prime\prime\prime}}^{{}}\right\rangle $, where the symbol
$\sum^{\prime}$ stands for summation over all $l^{\prime},l^{\prime\prime
},l^{\prime\prime\prime},l^{\prime\prime\prime\prime}\in\left\{
1,2,\cdots,L\right\}  $ that satisfy the momentum conservation condition
$l^{\prime}+l^{\prime\prime\prime}=l^{\prime\prime}+l^{\prime\prime
\prime\prime}$.

For the case $L=2$ (i.e. two sites), and for the subspace of two Bosons (i.e.
$\left\langle N\right\rangle =2$), the matrix representation of $\mathcal{H}%
_{\mathrm{B}}$ in the basis $\left\{  \left\vert n_{2}=0,n_{1}=2\right\rangle
,\left\vert n_{2}=1,n_{1}=1\right\rangle ,\left\vert n_{2}=2,n_{1}%
=0\right\rangle \right\}  $ is given by%
\begin{equation}
\mathcal{H}_{\mathrm{B}}\dot{=}U\left(
\begin{array}
[c]{ccc}%
1 & -\tau & 0\\
-\tau & 0 & -\tau\\
0 & -\tau & 1
\end{array}
\right)  \ ,\label{H TS BHM}%
\end{equation}
where $\tau=2^{3/2}t/U$. The $3\times3$ Hamiltonian matrix $\mathcal{H}%
_{\mathrm{B}}$ (\ref{H TS BHM}) is diagonalized by the unitary matrix $u$,
which is given by%
\begin{equation}
u=\left(
\begin{array}
[c]{ccc}%
\frac{\sin\frac{\alpha}{2}}{\sqrt{2}} & \frac{1}{\sqrt{2}} & \frac{\cos
\frac{\alpha}{2}}{\sqrt{2}}\\
-\cos\frac{\alpha}{2} & 0 & \sin\frac{\alpha}{2}\\
\frac{\sin\frac{\alpha}{2}}{\sqrt{2}} & -\frac{1}{\sqrt{2}} & \frac{\cos
\frac{\alpha}{2}}{\sqrt{2}}%
\end{array}
\right)  \ ,\label{u TS BHM}%
\end{equation}
where $\tan\alpha=-2^{3/2}\tau$. The energy eigenvalues $E_{n}$ are given by
$2E_{1}/U=1-\sqrt{1+\left(  8t/U\right)  ^{2}}$, $E_{2}=U$, and $2E_{3}%
/U=1+\sqrt{1+\left(  8t/U\right)  ^{2}}$. For a density operator $\rho$ having
a $3\times3$ matrix representation given by%
\begin{equation}
\rho\dot{=}\left(
\begin{array}
[c]{ccc}%
\cos^{2}\left(  x+\frac{\pi}{4}\right)  \cos^{2}\phi & \frac{a^{{}}}{2} &
\frac{c^{{}}}{2}\\
\frac{a^{\ast}}{2} & \sin^{2}\phi & \frac{b^{{}}}{2}\\
\frac{c^{\ast}}{2} & \frac{b^{\ast}}{2} & \sin^{2}\left(  x+\frac{\pi}%
{4}\right)  \cos^{2}\phi
\end{array}
\right)  \ ,
\end{equation}
where $x$ and $\phi$ are real, and $a$, $b$ and $c$ are complex, the energy
expectation value $\left\langle \mathcal{H}_{\mathrm{B}}\right\rangle $ is
given by $\left\langle \mathcal{H}_{\mathrm{B}}\right\rangle /U=\cos^{2}%
\phi-\tau\operatorname{Re}\left(  a+b\right)  $, and the following holds $\left\langle Q_{1,1}\right\rangle =\left\langle Q_{2,2}%
\right\rangle =-\left\langle Q_{1,2}\right\rangle =\left(  1-\sin^{2}\left(
2x\right)  \cos^{2}\phi\right)  \cos^{2}\phi$ [see Eq. (\ref{Q_jp,jpp})].

\begin{figure}[ptb]
\begin{center}
\includegraphics[width=3.2in,keepaspectratio]{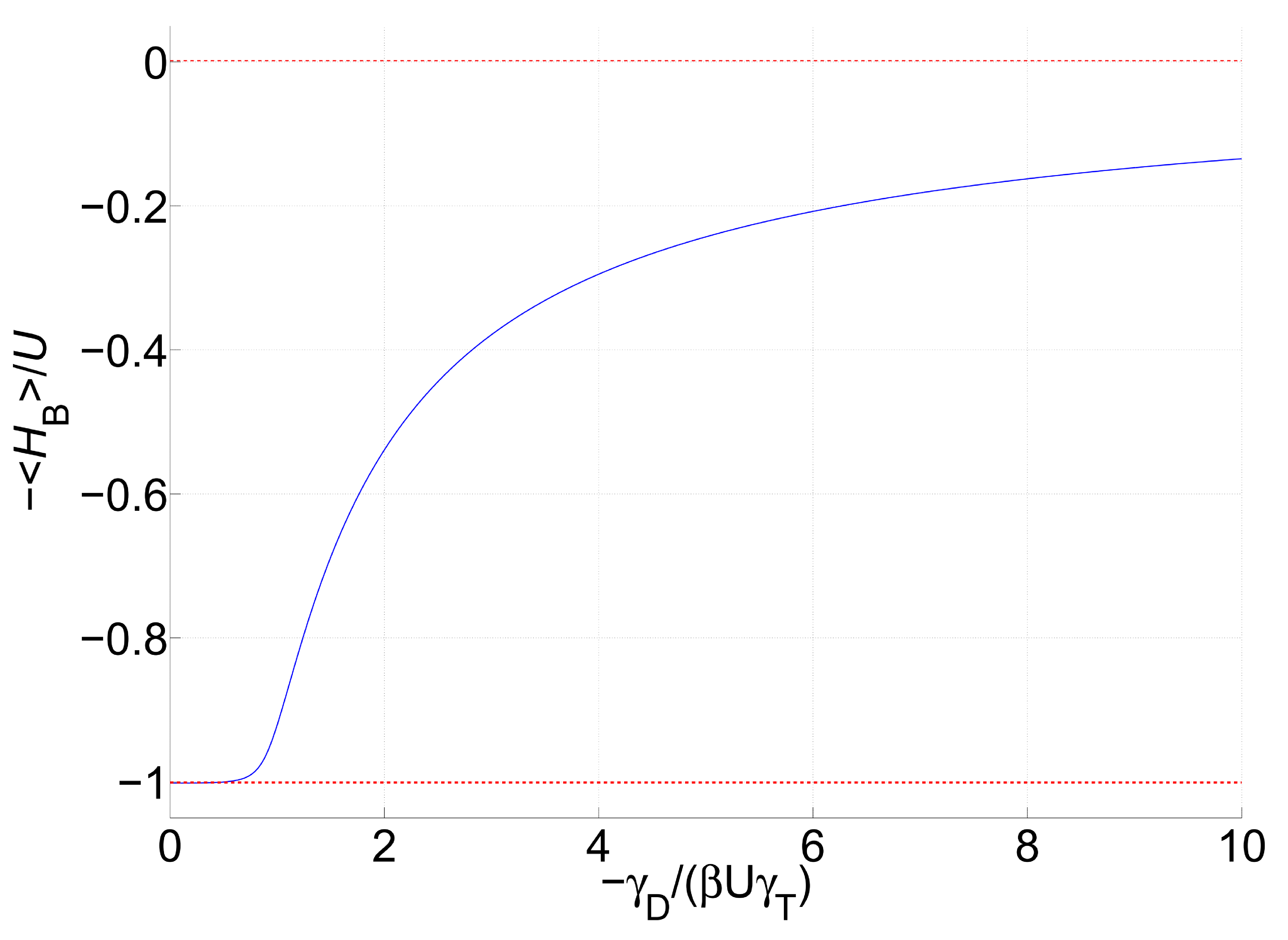}
\end{center}
\caption{{}Bose-Hubbard model. The steady state value of $\left\langle
\mathcal{H}_{\mathrm{B}}\right\rangle $ is found by numerically integrating
the master equation (\ref{MME}). The values $U=-1$ and $t/U=-10^{-2}$ are
assumed. Note that the gap between the ground state $E_{3}$ and
the first excited state $E_{2}$ is nearly invisible in the plot, since
$\left\vert t/U\right\vert \ll1$.}%
\label{FigBHM}%
\end{figure}

For attractive interaction (i.e. for $U<0$), the ground state has energy
$E_{3}$ [the corresponding state vector is given by the third column of the
unitary matrix $u$ given by Eq. (\ref{u TS BHM})]. In the absence of
disentanglement (i.e. for $\gamma_{\mathrm{D}}=0$), and in the low temperature
limit, the effective free energy $\left\langle \mathcal{U}_{\mathrm{e}%
}\right\rangle $ is locally minimized at the ground state. A stability
analysis yields that, for the case $\left\vert t/U\right\vert \ll1$, a
symmetry-breaking quantum phase transition occurs at $-\gamma_{\mathrm{D}%
}/\left(  \beta U\gamma_{\mathrm{T}}\right)  \simeq1/2$. This analytical
result is validated by numerically integrating the master equation
(\ref{MME}). The result is shown in Fig. \ref{FigBHM}, which depicts the
dependency of $\left\langle \mathcal{H}_{\mathrm{B}}\right\rangle $ in steady
state on $-\gamma_{\mathrm{D}}/\left(  \beta U\gamma_{\mathrm{T}}\right)  $.
The overlaid red dashed lines in Fig. \ref{FigBHM} indicate the eigenvalues
$E_{1}$, $E_{2}$ and $E_{3}$.

\textbf{Fermi-Hubbard model} - For the Fermi-Hubbard model, the array is
occupied by spin 1/2 Fermions. The Fermionic creation and annihilation
operators corresponding to site $l\in\left\{  1,2,\cdots,L\right\}  $ with
spin state $\sigma\in\left\{  \uparrow,\downarrow\right\}  $ are denoted by
$b_{l,\sigma}^{\dag}$ and $b_{l,\sigma}^{{}}$, respectively. The Hamiltonian
$\mathcal{H}_{\mathrm{F}}$ is given by%
\begin{align}
\mathcal{H}_{\mathrm{F}}  &  =-t\sum_{\sigma\in\left\{  \uparrow
,\downarrow\right\}  }\sum_{l=1}^{L}\left(  b_{l,\sigma}^{\dag}b_{l+1,\sigma
}^{{}}+b_{l+1,\sigma}^{\dag}b_{l,\sigma}^{{}}\right) \nonumber\\
&  +U\sum_{l=1}^{L}\left(  N_{l,\uparrow}-\frac{1}{2}\right)  \left(
N_{l,\downarrow}-\frac{1}{2}\right)  \ ,\nonumber\\
&  \label{H FHM}%
\end{align}
where $N_{l,\sigma}=b_{l,\sigma}^{\dag}b_{l,\sigma}^{{}}$. It is assumed that
$U>0$ (i.e. interaction is repulsive). For $\sigma\in\left\{  \uparrow
,\downarrow\right\}  $,\ the total spin $\sigma$\ number operator $N_{\sigma}%
$, which is defined by $N_{\sigma}=\sum_{l=1}^{L}N_{l,\sigma}$, is a constant
of the motion.

\begin{figure}[ptb]
\begin{center}
\includegraphics[width=3.2in,keepaspectratio]{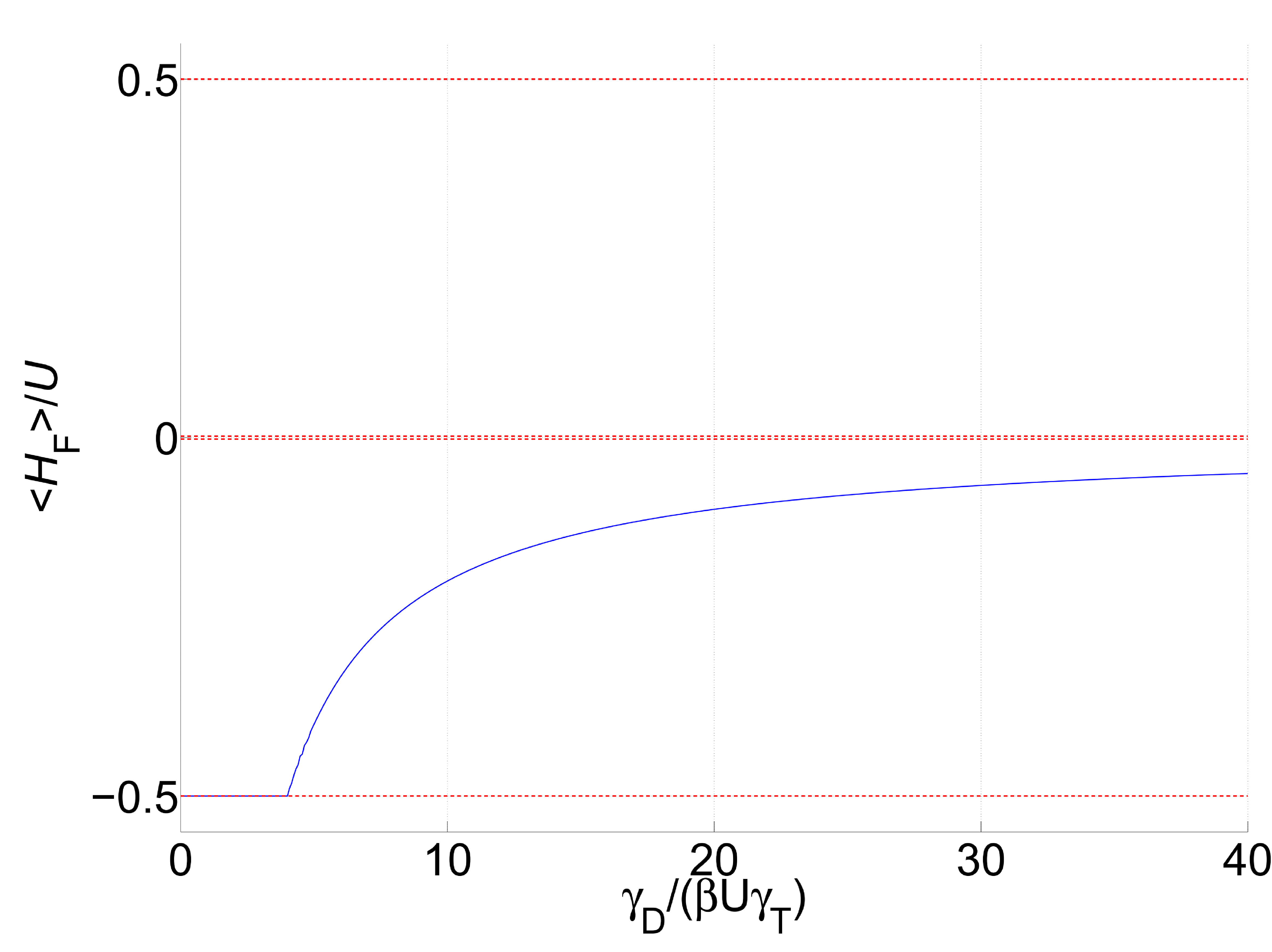}
\end{center}
\caption{{}Fermi-Hubbard model. The values $U=1$ and $t/U=10^{-3}$ are assumed
for the numerical integrating of the master equation (\ref{MME}). The overlaid red dashed horizontal lines near the values $-0.5$, $0$ and $0.5$
of the ratio $\left\langle \mathcal{H}_{\mathrm{F}}\right\rangle /U $
represent groups containing, respectively, $4$, $8$ and $4$ nearly degenerate
energy eigenvalues.}%
\label{FigFHM}%
\end{figure}

For the case $L=2$, the floor (lowest energy) state $\left\vert f\right\rangle
$ and the ceiling (highest energy) state $\left\vert c\right\rangle $ of the
Hamiltonian $\mathcal{H}_{\mathrm{F}}$ (\ref{H FHM}) are given by%
\begin{align}
\left\vert f\right\rangle  &  =\frac{\left(  \left\vert 0011\right\rangle
+\left\vert 1100\right\rangle \right)  \cos\alpha+\left(  \left\vert
0110\right\rangle +\left\vert 1001\right\rangle \right)  \sin\alpha}{\sqrt{2}%
}\ ,\\
\left\vert c\right\rangle  &  =\frac{\left(  \left\vert 0011\right\rangle
+\left\vert 1100\right\rangle \right)  \sin\alpha-\left(  \left\vert
0110\right\rangle +\left\vert 1001\right\rangle \right)  \cos\alpha}{\sqrt{2}%
}\ ,
\end{align}
where $\alpha=\left(  1/2\right)  \tan^{-1}\left(  -8t/U\right)  $. The energy
expectation value $\left\langle \mathcal{H}_{\mathrm{F}}\right\rangle $, which
is calculated by numerically integrating the master equation (\ref{MME}), is
plotted in Fig. \ref{FigFHM} as a function $\gamma_{\mathrm{D}}/\left(  \beta
U\gamma_{\mathrm{T}}\right)  $. Both below and above the phase transition seen
in Fig. \ref{FigFHM}, the occupation probabilities of all states, except of
the floor $\left\vert f\right\rangle $ and the ceiling $\left\vert
c\right\rangle $ ones, is found to be small (below $10^{-5}$ for the range
plotted in Fig. \ref{FigFHM}).

Consider a pure normalized state $\left\vert \psi\right\rangle $ given by
$\left\vert \psi\right\rangle =e^{i\varphi}\cos\left(  \phi\right)  \left\vert
f\right\rangle +e^{-i\varphi}\sin\left(  \phi\right)  \left\vert
c\right\rangle $, where both $\phi$ and $\varphi\ $are real. The following
holds $\left\langle \psi\right\vert \mathcal{H}_{\mathrm{F}}\left\vert
\psi\right\rangle /U=-\cos\left(  2\phi\right)  /\cos\left(  2\alpha\right)  $
and $\left\langle N_{1,\uparrow}N_{1,\downarrow}\right\rangle -\left\langle
N_{1,\uparrow}\right\rangle \left\langle N_{1,\downarrow}\right\rangle
=\left\langle N_{2,\uparrow}N_{2,\downarrow}\right\rangle -\left\langle
N_{2,\uparrow}\right\rangle \left\langle N_{2,\downarrow}\right\rangle
\equiv\nu$, where $\nu=\left(  1/4\right)  \cos\left(  2\alpha\right)  \left(
\tan\left(  2\alpha\right)  \sin\left(  2\phi\right)  \cos\left(
2\varphi\right)  -\cos\left(  2\phi\right)  \right)  $. In the absence of
disentanglement (i.e. for $\gamma_{\mathrm{D}}=0$), and in the low temperature
limit, the effective free energy $\left\langle \mathcal{U}_{\mathrm{e}%
}\right\rangle $ is locally minimized at the floor (ground) state $\left\vert
f\right\rangle $. A stability analysis for the case $\left\vert t/U\right\vert
\ll1$ yields a symmetry-breaking quantum phase transition occurring at
$\gamma_{\mathrm{D}}/\left(  \beta U\gamma_{\mathrm{T}}\right)  =4$. This
analytical result is validated by the plot shown in Fig. \ref{FigFHM}. The
overlaid red dashed lines in Fig. \ref{FigFHM} indicate the $16$ energy
eigenvalues of $\mathcal{H}_{\mathrm{F}}$.

\textbf{Discussion} - Below the phase transition (see Fig.
\ref{FigBHM} for Bosons, and Fig. \ref{FigFHM} for Fermions), in steady state
the ground state is nearly fully occupied (in the low-temperature limit). The
disentanglement term added to the master equation (\ref{MME}) suppresses the
expectation value $\left\langle \mathcal{Q}^{\left(  \mathrm{D}\right)
}\right\rangle $, which quantifies correlation between particles. For both
above-discussed examples, the value of $\left\langle \mathcal{Q}^{\left(
\mathrm{D}\right)  }\right\rangle $ is not minimized at the ground state.
Consequently, when the rate of disentanglement $\gamma_{\mathrm{D}}$ is
sufficiently large, occupying higher energy states becomes preferable, since
that allows lowering the value of $\left\langle \mathcal{Q}^{\left(
\mathrm{D}\right)  }\right\rangle $.

The modified mater equation given by Eq. (\ref{MME}) is by no means unique.
Alternative mechanisms giving rise to disentanglement can be explored by
redefining the operator $\mathcal{Q}^{\left(  \mathrm{D}\right)  }$. A variety
of methods have been proposed to quantify entanglement of indistinguishable
particles
\cite{Benatti_1,Zanardi_042101,Schliemann_022303,Li_054302,Kent_1839,Pearle_2277}%
. Some of these proposals can yield alternative definitions for the
disentanglement operator $\mathcal{Q}^{\left(  \mathrm{D}\right)  }$. In the
current study, the definition of the operator $\mathcal{Q}^{\left(
\mathrm{D}\right)  }$ is based on two-particle interaction [see Eq.
(\ref{Q_jp,jpp})]. This chosen definition of $\mathcal{Q}^{\left(
\mathrm{D}\right)  }$\ allows making disentanglement local (since
disentanglement becomes active only when particles interact). In contrast,
alternative definitions that enable remote disentanglement might give rise to
a conflict with the causality principle
\cite{Bassi_055027,Jordan_022101,Polchinski_397,Helou_012021,Rembielinski_012027,Rembielinski_420}.

The spontaneous disentanglement hypothesis is inherently falsifiable, because
it yields predictions, which are experimentally distinguishable from
predictions obtained from standard QM. For finite $L$, standard QM does not
yield any phase transition in both Bose-Hubbard and Fermi-Hubbard models. On
the other hand, phase transitions become possible when the mean field approximation is employed. For example, for the
Fermi-Hubbard model, the mean field approximation
yields a phase transition that occurs provided that the Stoner criterion is
satisfied \cite{Stoner_372}. However, it has remained unclear how the mean field approximation, which is based on the
assumption that entanglement can be disregarded, can be justified within the
framework of standard QM. On the other hand, as is demonstrated by Figs.
\ref{FigBHM} and \ref{FigFHM} for both Bosons and Fermions, in the presence of
spontaneous disentanglement quantum phase transitions can occur in systems
containing a finite number of particles.

\textbf{Summary} - Disentanglement of indistinguishable particles is explored.
It is found for both Bosons and Fermions that disentanglement can give rise to
quantum phase transitions, provided that the rate of disentanglement
$\gamma_{\mathrm{D}}$ is sufficiently large. Further study is needed to
determine whether the hypothesis that spontaneous disentanglement occurs in
quantum systems is consistent with experimental observations \cite{Roch_633,Thomas_145,Trishin_236801,Blesio_045113}.

\bibliographystyle{ieeepes}
\bibliography{acompat,Eyal_Bib}

\newif\ifabfull\abfulltrue
\begin{thebibliography}{10}

\bibitem{Lindblad_119}
Goran Lindblad,
\newblock ``On the generators of quantum dynamical semigroups'',
\newblock {\em Communications in Mathematical Physics}, vol. 48, no. 2, pp.
  119--130, 1976.

\bibitem{Weinberg_61}
Steven Weinberg,
\newblock ``Precision tests of quantum mechanics'',
\newblock in {\em THE OSKAR KLEIN MEMORIAL LECTURES 1988--1999}, pp. 61--68.
  World Scientific, 2014.

\bibitem{Doebner_3764}
H-D Doebner and Gerald~A Goldin,
\newblock ``Introducing nonlinear gauge transformations in a family of
  nonlinear schr{\"o}dinger equations'',
\newblock {\em Physical Review A}, vol. 54, no. 5, pp. 3764, 1996.

\bibitem{Gisin_5677}
Nicolas Gisin and Ian~C Percival,
\newblock ``The quantum-state diffusion model applied to open systems'',
\newblock {\em Journal of Physics A: Mathematical and General}, vol. 25, no.
  21, pp. 5677, 1992.

\bibitem{Gisin_2259}
Nicolas Gisin,
\newblock ``A simple nonlinear dissipative quantum evolution equation'',
\newblock {\em Journal of Physics A: Mathematical and General}, vol. 14, no. 9,
  pp. 2259, 1981.

\bibitem{Kaplan_055002}
David~E Kaplan and Surjeet Rajendran,
\newblock ``Causal framework for nonlinear quantum mechanics'',
\newblock {\em Physical Review D}, vol. 105, no. 5, pp. 055002, 2022.

\bibitem{Munoz_110503}
Manuel~H Mu{\~n}oz-Arias, Pablo~M Poggi, Poul~S Jessen, and Ivan~H Deutsch,
\newblock ``Simulating nonlinear dynamics of collective spins via quantum
  measurement and feedback'',
\newblock {\em Physical review letters}, vol. 124, no. 11, pp. 110503, 2020.

\bibitem{Jacobs_279}
Kurt Jacobs and Daniel~A Steck,
\newblock ``A straightforward introduction to continuous quantum measurement'',
\newblock {\em Contemporary Physics}, vol. 47, no. 5, pp. 279--303, 2006.

\bibitem{Geller_2200156}
Michael~R Geller,
\newblock ``Fast quantum state discrimination with nonlinear positive
  trace-preserving channels'',
\newblock {\em Advanced Quantum Technologies}, p. 2200156, 2023.

\bibitem{Schrodinger_807}
E.~Schrodinger,
\newblock ``Die gegenwartige situation in der quantenmechanik'',
\newblock {\em Naturwissenschaften}, vol. 23, pp. 807, 1935.

\bibitem{Bassi_471}
Angelo Bassi, Kinjalk Lochan, Seema Satin, Tejinder~P Singh, and Hendrik
  Ulbricht,
\newblock ``Models of wave-function collapse, underlying theories, and
  experimental tests'',
\newblock {\em Reviews of Modern Physics}, vol. 85, no. 2, pp. 471, 2013.

\bibitem{Pearle_857}
Philip Pearle,
\newblock ``Reduction of the state vector by a nonlinear schr{\"o}dinger
  equation'',
\newblock {\em Physical Review D}, vol. 13, no. 4, pp. 857, 1976.

\bibitem{Ghirardi_470}
Gian~Carlo Ghirardi, Alberto Rimini, and Tullio Weber,
\newblock ``Unified dynamics for microscopic and macroscopic systems'',
\newblock {\em Physical review D}, vol. 34, no. 2, pp. 470, 1986.

\bibitem{Bassi_257}
Angelo Bassi and GianCarlo Ghirardi,
\newblock ``Dynamical reduction models'',
\newblock {\em Physics Reports}, vol. 379, no. 5-6, pp. 257--426, 2003.

\bibitem{Bennett_170502}
Charles~H Bennett, Debbie Leung, Graeme Smith, and John~A Smolin,
\newblock ``Can closed timelike curves or nonlinear quantum mechanics improve
  quantum state discrimination or help solve hard problems?'',
\newblock {\em Physical review letters}, vol. 103, no. 17, pp. 170502, 2009.

\bibitem{Kowalski_1}
Krzysztof Kowalski,
\newblock ``Linear and integrable nonlinear evolution of the qutrit'',
\newblock {\em Quantum Information Processing}, vol. 19, no. 5, pp. 1--31,
  2020.

\bibitem{Fernengel_385701}
Bernd Fernengel and Barbara Drossel,
\newblock ``Bifurcations and chaos in nonlinear lindblad equations'',
\newblock {\em Journal of Physics A: Mathematical and Theoretical}, vol. 53,
  no. 38, pp. 385701, 2020.

\bibitem{Kowalski_167955}
K~Kowalski and J~Rembieli{\'n}ski,
\newblock ``Integrable nonlinear evolution of the qubit'',
\newblock {\em Annals of Physics}, vol. 411, pp. 167955, 2019.

\bibitem{Oppenheim_041040}
Jonathan Oppenheim,
\newblock ``A postquantum theory of classical gravity?'',
\newblock {\em Physical Review X}, vol. 13, no. 4, pp. 041040, 2023.

\bibitem{Schrinski_133604}
Bj{\"o}rn Schrinski, Yu~Yang, Uwe von L{\"u}pke, Marius Bild, Yiwen Chu, Klaus
  Hornberger, Stefan Nimmrichter, and Matteo Fadel,
\newblock ``Macroscopic quantum test with bulk acoustic wave resonators'',
\newblock {\em Physical Review Letters}, vol. 130, no. 13, pp. 133604, 2023.

\bibitem{Chomaz_68}
Philippe Chomaz and Francesca Gulminelli,
\newblock ``Phase transitions in finite systems'',
\newblock in {\em Dynamics and thermodynamics of systems with long-range
  interactions}, pp. 68--129. Springer, 2002.

\bibitem{mainwood2005phase}
Paul Mainwood,
\newblock ``Phase transitions in finite systems'',
\newblock 2005.

\bibitem{Callender_539}
Craig Callender,
\newblock ``Taking thermodynamics too seriously'',
\newblock {\em Studies in history and philosophy of science part B: studies in
  history and philosophy of modern physics}, vol. 32, no. 4, pp. 539--553,
  2001.

\bibitem{Liu_S92}
Chuang Liu,
\newblock ``Explaining the emergence of cooperative phenomena'',
\newblock {\em Philosophy of Science}, vol. 66, no. S3, pp. S92--S106, 1999.

\bibitem{Ardourel_99}
Vincent Ardourel and Sorin Bangu,
\newblock ``Finite-size scaling theory: Quantitative and qualitative approaches
  to critical phenomena'',
\newblock {\em Studies in History and Philosophy of Science}, vol. 100, pp.
  99--106, 2023.

\bibitem{Shech_1170}
Elay Shech,
\newblock ``What is the paradox of phase transitions?'',
\newblock {\em Philosophy of Science}, vol. 80, no. 5, pp. 1170--1181, 2013.

\bibitem{toda1978statistical_I}
Morikazu Toda, Ryogo Kubo, Nobuhiko Sait{\=o}, Natsuki Hashitsume, and Natsuki
  Hashitsume,
\newblock {\em Statistical physics I},
\newblock Springer Science, 1978.

\bibitem{Sakthivadivel_035}
Dalton~AR Sakthivadivel,
\newblock ``Magnetisation and mean field theory in the ising model'',
\newblock {\em SciPost Physics Lecture Notes}, p. 035, 2022.

\bibitem{Suhl_209}
H~Suhl,
\newblock ``The theory of ferromagnetic resonance at high signal powers'',
\newblock {\em Journal of Physics and Chemistry of Solids}, vol. 1, no. 4, pp.
  209--227, 1957.

\bibitem{Buks_012439}
Eyal Buks,
\newblock ``Disentanglement-induced multistability'',
\newblock {\em Physical Review A}, vol. 110, no. 1, pp. 012439, 2024.

\bibitem{Bassi_055027}
Angelo Bassi and Kasra Hejazi,
\newblock ``No-faster-than-light-signaling implies linear evolution. a
  re-derivation'',
\newblock {\em European Journal of Physics}, vol. 36, no. 5, pp. 055027, 2015.

\bibitem{Jordan_022101}
Thomas~F. Jordan,
\newblock ``Assumptions that imply quantum dynamics is linear'',
\newblock {\em Phys. Rev. A}, vol. 73, pp. 022101, Feb 2006.

\bibitem{Polchinski_397}
Joseph Polchinski,
\newblock ``Weinberg?s nonlinear quantum mechanics and the
  einstein-podolsky-rosen paradox'',
\newblock {\em Physical Review Letters}, vol. 66, no. 4, pp. 397, 1991.

\bibitem{Helou_012021}
Bassam Helou and Yanbei Chen,
\newblock ``Extensions of born?s rule to non-linear quantum mechanics, some of
  which do not imply superluminal communication'',
\newblock in {\em Journal of Physics: Conference Series}. IOP Publishing, 2017,
  vol. 880, p. 012021.

\bibitem{Rembielinski_012027}
Jakub Rembieli{\'n}ski and Pawe{\l} Caban,
\newblock ``Nonlinear evolution and signaling'',
\newblock {\em Physical Review Research}, vol. 2, no. 1, pp. 012027, 2020.

\bibitem{Rembielinski_420}
Jakub Rembieli{\'n}ski and Pawe{\l} Caban,
\newblock ``Nonlinear extension of the quantum dynamical semigroup'',
\newblock {\em Quantum}, vol. 5, pp. 420, 2021.

\bibitem{Hejlesen_thesis}
Zakarias~Laberg Hejlesen,
\newblock ``Nonlinear quantum mechanics'',
\newblock Master's thesis, 2019.

\bibitem{Jordan_012010}
Thomas~F Jordan,
\newblock ``Why quantum dynamics is linear'',
\newblock in {\em Journal of Physics: Conference Series}. IOP Publishing, 2009,
  vol. 196, p. 012010.

\bibitem{Buks_2400036}
Eyal Buks,
\newblock ``Spontaneous disentanglement and thermalization'',
\newblock {\em Advanced Quantum Technologies}, p. 2400036, 2024.

\bibitem{Hubbard_238}
John Hubbard,
\newblock ``Electron correlations in narrow energy bands'',
\newblock {\em Proceedings of the Royal Society of London. Series A.
  Mathematical and Physical Sciences}, vol. 276, no. 1365, pp. 238--257, 1963.

\bibitem{Hubbard_401}
John Hubbard,
\newblock ``Electron correlations in narrow energy bands iii. an improved
  solution'',
\newblock {\em Proceedings of the Royal Society of London. Series A.
  Mathematical and Physical Sciences}, vol. 281, no. 1386, pp. 401--419, 1964.

\bibitem{Gutzwiller_159}
Martin~C Gutzwiller,
\newblock ``Effect of correlation on the ferromagnetism of transition metals'',
\newblock {\em Physical Review Letters}, vol. 10, no. 5, pp. 159, 1963.

\bibitem{Kanamori_275}
Junjiro Kanamori,
\newblock ``Electron correlation and ferromagnetism of transition metals'',
\newblock {\em Progress of Theoretical Physics}, vol. 30, no. 3, pp. 275--289,
  1963.

\bibitem{fetter2012quantum}
Alexander~L Fetter and John~Dirk Walecka,
\newblock {\em Quantum theory of many-particle systems},
\newblock Courier Corporation, 2012.

\bibitem{Grimaudo_033835}
R~Grimaudo, Asm De~Castro, M~Ku{\'s}, and A~Messina,
\newblock ``Exactly solvable time-dependent pseudo-hermitian su (1, 1)
  hamiltonian models'',
\newblock {\em Physical Review A}, vol. 98, no. 3, pp. 033835, 2018.

\bibitem{Grabert_161}
H~Grabert,
\newblock ``Nonlinear relaxation and fluctuations of damped quantum systems'',
\newblock {\em Zeitschrift f{\"u}r Physik B Condensed Matter}, vol. 49, no. 2,
  pp. 161--172, 1982.

\bibitem{Ottinger_052119}
Hans~Christian {\"O}ttinger,
\newblock ``Nonlinear thermodynamic quantum master equation: Properties and
  examples'',
\newblock {\em Physical Review A}, vol. 82, no. 5, pp. 052119, 2010.

\bibitem{Jaynes_579}
Edwin~T Jaynes,
\newblock ``The minimum entropy production principle'',
\newblock {\em Annual Review of Physical Chemistry}, vol. 31, no. 1, pp.
  579--601, 1980.

\bibitem{Buks_052217}
Eyal Buks and Dvir Schwartz,
\newblock ``Stability of the grabert master equation'',
\newblock {\em Physical Review A}, vol. 103, no. 5, pp. 052217, 2021.

\bibitem{Lieb_1}
Elliott~H Lieb and FY~Wu,
\newblock ``The one-dimensional hubbard model: a reminiscence'',
\newblock {\em Physica A: statistical mechanics and its applications}, vol.
  321, no. 1-2, pp. 1--27, 2003.

\bibitem{Links_1591}
Jon Links, Angela Foerster, Arlei~Prestes Tonel, and Gilberto Santos,
\newblock ``The two-site bose--hubbard model'',
\newblock in {\em Annales Henri Poincar{\'e}}. Springer, 2006, vol.~7, pp.
  1591--1600.

\bibitem{Zhang_591_}
JM~Zhang and RX~Dong,
\newblock ``Exact diagonalization: the bose--hubbard model as an example'',
\newblock {\em European Journal of Physics}, vol. 31, no. 3, pp. 591, 2010.

\bibitem{Amico_4298}
Luigi Amico, Davit Aghamalyan, Filip Auksztol, Herbert Crepaz, Rainer Dumke,
  and Leong~Chuan Kwek,
\newblock ``Superfluid qubit systems with ring shaped optical lattices'',
\newblock {\em Scientific reports}, vol. 4, no. 1, pp. 4298, 2014.

\bibitem{Benatti_1}
Fabio Benatti, Roberto Floreanini, Fabio Franchini, and Ugo Marzolino,
\newblock ``Entanglement in indistinguishable particle systems'',
\newblock {\em Physics Reports}, vol. 878, pp. 1--27, 2020.

\bibitem{Zanardi_042101}
Paolo Zanardi,
\newblock ``Quantum entanglement in fermionic lattices'',
\newblock {\em Physical review A}, vol. 65, no. 4, pp. 042101, 2002.

\bibitem{Schliemann_022303}
John Schliemann, J~Ignacio Cirac, Marek Ku{\'s}, Maciej Lewenstein, and Daniel
  Loss,
\newblock ``Quantum correlations in two-fermion systems'',
\newblock {\em Physical Review A}, vol. 64, no. 2, pp. 022303, 2001.

\bibitem{Li_054302}
Yan~Song Li, Bei Zeng, Xiao~Shu Liu, and Gui~Lu Long,
\newblock ``Entanglement in a two-identical-particle system'',
\newblock {\em Physical Review A}, vol. 64, no. 5, pp. 054302, 2001.

\bibitem{Kent_1839}
Adrian Kent,
\newblock ``?quantum jumps? and indistinguishability'',
\newblock {\em Modern Physics Letters A}, vol. 4, no. 19, pp. 1839--1845, 1989.

\bibitem{Pearle_2277}
Philip Pearle,
\newblock ``Combining stochastic dynamical state-vector reduction with
  spontaneous localization'',
\newblock {\em Phys. Rev. A}, vol. 39, pp. 2277--2289, 1989.

\bibitem{Stoner_372}
Edmund~Clifton Stoner,
\newblock ``Collective electron ferromagnetism'',
\newblock {\em Proceedings of the Royal Society of London. Series A.
  Mathematical and Physical Sciences}, vol. 165, no. 922, pp. 372--414, 1938.

\bibitem{Roch_633}
Nicolas Roch, Serge Florens, Vincent Bouchiat, Wolfgang Wernsdorfer, and Franck
  Balestro,
\newblock ``Quantum phase transition in a single-molecule quantum dot'',
\newblock {\em Nature}, vol. 453, no. 7195, pp. 633--637, 2008.

\bibitem{Thomas_145}
L~Thomas, FL~Lionti, R~Ballou, Dante Gatteschi, Roberta Sessoli, and B~Barbara,
\newblock ``Macroscopic quantum tunnelling of magnetization in a single crystal
  of nanomagnets'',
\newblock {\em Nature}, vol. 383, no. 6596, pp. 145--147, 1996.

\bibitem{Trishin_236801}
Sergey Trishin, Christian Lotze, Nils Bogdanoff, Felix von Oppen, and
  Katharina~J Franke,
\newblock ``Moir{\'e} tuning of spin excitations: Individual fe atoms on mos
  2/au (111)'',
\newblock {\em Physical Review Letters}, vol. 127, no. 23, pp. 236801, 2021.

\bibitem{Blesio_045113}
GG~Blesio and AA~Aligia,
\newblock ``Topological quantum phase transition in individual fe atoms on mos
  2/au (111)'',
\newblock {\em Physical Review B}, vol. 108, no. 4, pp. 045113, 2023.

\end{thebibliography}

\end{document}